\documentclass{elsart}

\usepackage{amssymb}
\usepackage{graphicx}

\newcommand{\quarterthin}{\kern 0.0417em}

\begin{document}

\begin{frontmatter}

\title{
Coexistence of normal, super-, and hyper-deformation in nuclei: A
study with angular momentum projection}

\author[1,2]{Yang Sun},
\author[3]{Jing-ye Zhang},
\author[4,5]{Gui-Lu Long},
\author[6]{Cheng-Li Wu}

\address[1]{Department of Physics, Shanghai Jiao Tong University,
Shanghai 200240, P. R. China}
\address[2]{Joint Institute for Nuclear Astrophysics,
University of Notre Dame, Notre Dame, Indiana 46556, USA}
\address[3]{Department of Physics and Astronomy, University of
Tennessee, Knoxville, Tennessee 37996, USA}
\address[4]{Department of Physics, Tsinghua University, Beijing
100084, P.R. China}
\address[5]{Center of Nuclear Theory, Lanzhou National
Laboratory of Heavy Ion Accelerator, Lanzhou 730000, P. R. China}
\address[6]{Department of Physics, Chung Yuan
Christian University, Chung-Li, Taiwan 32023, ROC}

\begin{abstract}
Angular-momentum-projected energy surface calculations for
$A\approx$ 110 nuclei indicate three distinct energy minima
occurring at different angular-momenta. These correspond to normal,
super-, and hyper-deformed shapes coexisting in one nucleus.
$^{110}$Pd is studied in detail, with a quantitative prediction on
super- and hyper-deformed spectra by the Projected Shell Model
calculation. It is suggested that several other neighboring nuclei
in the $A$-110 mass region, with the neutron number around 64, also
exhibit clear super- and hyper-deformation minima.
\end{abstract}

\end{frontmatter}

\newcommand{\epsfigboxL}[5]{%
\begin{figure} \vspace{#3}%
\includegraphics[width=12cm]{#2}%
\caption{ \label{fig:#1} #5} \vspace{#4}
\end{figure}}

\newcommand{\epsfigboxM}[5]{%
\begin{figure} \vspace{#3}%
\includegraphics[width=10cm]{#2}%
\caption{ \label{fig:#1} #5} \vspace{#4}
\end{figure}}

\newcommand{\epsfigboxS}[5]{%
\begin{figure} \vspace{#3}%
\includegraphics[width=8.5cm]{#2}%
\caption{ \label{fig:#1} #5} \vspace{#4}
\end{figure}}


\newcommand{\epstblbox}[6]{%
\begin{table}\vspace{#3}%
\caption{ \label{tab:#1}#5}
\includegraphics{#2}
\begin{flushleft}\vspace*{-4pt}#6
\end{flushleft}
\vspace{#4}
\end{table}}

\section{Introduction}

The usual assumption of nuclear structure models is that a nucleus
can be described as a set of interacting particles moving in an
average potential. The nucleonic motion in a potential well often
occurs in a highly coherent way, giving rise to collective phenomena
such as rotational spectra and large quadrupole moments. These led
to the idea of introduction of the concept of nuclear deformation
\cite{BMbook}. Since then, the search for exotic nuclear shapes has
continuously been one of the research frontiers in nuclear structure
physics. One remarkable example is the discovery of super-deformed
(SD) rotational bands \cite{Twin88}. Moreover, microscopic
calculations \cite{Dudek88,WD95,Ch01,BMP02} have predicted that some
nuclei may attain even greater quadrupole deformations corresponding
to a hyper-deformed (HD) shape, and that the nuclei with these
deformations are possibly stable against nuclear fission.
Experimentally, the search for HD states has been continued for
years (see, for example, Ref. \cite{HD} and conference proceedings
of the 2004 Zakopane School \cite{Polon}).

The observation of rotational bands in $^{108}$Cd has stimulated a
new wave of study along this line. In Ref. \cite{Exp}, Clark {\it et
al.} reported a rotational band with the data indicating that this
nucleus has the most deformed shape identified to date. Using the
Projected Shell Model (PSM) \cite{PSM}, some of us analyzed the
structure of the band. It was found \cite{Lee02} that the
wavefunctions of the observed states notably contain a component of
the proton $i_{13/2}$ orbital from the $N=6$ harmonic oscillator
shell. It is well-known that the $\pi i_{13/2}$ orbitals usually
appear near the proton Fermi level in normal-deformed (ND) nuclei of
the actinide region, and near those in SD nuclei of the rare-earth
region. In a $Z=48$ nucleus of the $A$-110 mass region, occupation
of the $\pi i_{13/2}$ orbitals is possible only if the nucleus has a
greater deformation.

Inspired by these advances, we have employed the PSM to carry out a
systematical study on the neighboring nuclei of $^{108}$Cd.
Angular-momentum-projected energy surface calculations have explored
distinct minima in some of these nuclei, separately lying at
different deformations and angular momenta. Based on the content of
the wavefunctions, in which nucleons from several major shells
contribute to the collectivity near the Fermi surface, we can
understand the occurrence of the minima as a correlation effect
involving multiple major shells. According to our analysis,
$^{108}$Cd has two energy minima. The observed band in $^{108}$Cd
\cite{Exp} (with neutron number 60) belongs to the second minimum.
As we shall show in the present paper, three energy minima, which
correspond to ND, SD, and HD shapes coexisting in one nucleus, can
be formed in nuclei with neutron number around 64.

The paper is organized as follows. In Section 2, we outline the
theory and give the calculation conditions. In Section 3, we take
$^{110}$Pd as an example to present the main results of the present
study. Further quantitative predictions are given in Section 4.
Finally, the paper is summarized in Section 5.

\section{Theory of angular momentum projection}

Using a conventional shell model to study deformed, heavy nuclei is
desirable. However, it is unfeasible in practice because of the
unavoidable problem related to huge basis dimensions. Therefore, the
study of deformed, heavy nuclei has relied mainly on the mean field
approximation. In a mean field theory, single particle motion is
treated properly which gives rise to a correct description of shell
structure. However, correlations beyond the mean field level are
usually neglected. If the energy gain due to the correlations would
be a constant with no particle-number and angular-momentum
dependence, then consideration of correlations may not be important.
However, it is often not the case in reality. As we shall see in the
$^{110}$Pd example, when the correlation effect is added to that of
shell structure, several energy minima clearly occur in the
projected energy curves.

The Projected Shell Model \cite{PSM} is a shell model built from a
deformed basis. To be concrete, the PSM constructs its shell-model
space by using the deformed Nilsson single-particle states with a
deformation $\varepsilon_2$. Pairing correlations are incorporated
into the Nilsson states by the BCS calculations. The consequence of
the Nilsson-BCS calculations defines a set of quasiparticle (qp)
states associated with the qp vacuum
$\left|\phi(\varepsilon_2)\right> \equiv \left|0\right>$ in the
intrinsic frame. One then builds shell model bases by writing
multi-qp states on top of $\left|0\right>$:
\begin{equation}
\left|\Phi_\kappa\right> = \left\{\left|0 \right>, \
\alpha^\dagger_{n_i} \alpha^\dagger_{n_j} \left|0 \right>,\
\alpha^\dagger_{p_i} \alpha^\dagger_{p_j} \left|0 \right>,\
\alpha^\dagger_{n_i} \alpha^\dagger_{n_j} \alpha^\dagger_{p_i}
\alpha^\dagger_{p_j} \left|0 \right>, \cdots \right\} ,
\label{baset}
\end{equation}
where $\alpha^\dagger$ is the qp creation operator and the index $n$
($p$) denotes neutron (proton) Nilsson quantum numbers of the
orbitals. In Eq. (\ref{baset}), when the higher order multi-qp
states denoted by ``$\cdots$" are all included, one recovers the
complete shell model configuration. In practice, these high order
multi-qp states are not necessarily needed if one is interested in
the near-Yrast property only. We have checked that for producing
energy minima reported in the present paper, nearly the same results
are obtained when the basis size in (\ref{baset}) is changed from a
few to a few hundreds. Angular momentum projection transforms the
wavefunction from the intrinsic frame to the laboratory frame. The
total wavefunction is expressed as
\begin{equation}
\left|\Psi^I_M\right> = \sum_\kappa f^I_\kappa \hat
P^I_{MK_\kappa} \left|\Phi_\kappa\right> , \label{wavef}
\end{equation}
where $\hat P^I$ is the angular-momentum-projection operator
\cite{RS80} and $\kappa$ labels the basis states. Finally a two-body
shell model Hamiltonian is diagonalized in the projected space, and
the diagonalization determines $f^I_\kappa$ in (\ref{wavef}). This
relatively-simple model has been extensively tested in the past
decade. For the applications related to the present work, the PSM
has been successful in describing the SD states in different mass
regions \cite{Sun95,Sun97,Sun99,Sun01}.

In searching for energy minima, we calculate the
angular-momentum-projected energy
\begin{equation}
E^I = {{\left< \Psi^I\right|\hat H \left| \Psi^I\right>} \over
{\left< \Psi^I | \Psi^I\right>}}
\label{energy}
\end{equation}
as a function of static quadrupole moment defined by
\begin{equation}
\left< Q \right> = e_{\rm eff}{{\left<\Psi^I ||\hat Q
||\Psi^I\right>} \over {\left< I020|I0 \right>}}, \label{Qs}
\end{equation}
where the effective charges are $e_{\rm eff}=1.5e$ for protons and
0.5$e$ for neutrons. The energy $E^I$ and the variable $\left< Q
\right>$ are both evaluated by using the same wavefunction
(\ref{wavef}).

Equation (3) represents a study of energy minimization after angular
momentum projection. In principle, the energies calculated with
$\left| \Psi^I\right>$ can have advantages over those with
unprojected state $\left|0\right>$ in that the projection treats the
states fully quantum-mechanically by collecting all the
energy-degenerate mean-field states associated with the rotational
symmetry. Angular momentum projection is a superposition in the
Euler space which lowers the total energy of the system. The energy
gain is due to the correlations, which is particularly important in
the present content for the formation of local deformation minima.
Although in some cases, especially when the minima are deep and
stable, the results between projected and unprojected calculations
are qualitatively similar, there have been many examples with
qualitatively different conclusions (as for instance, the one shown
in the seminal work of Hayashi, Hara, and Ring \cite{HHR84}). In
recent years, there have been attempts to incorporate
angular-momentum-projection in the theories with effective
interactions \cite{Egido93,Egido04,Ring06,Tanabe00,Valor00}, and
notable effects due to projection have been found.

The Hamiltonian appearing in (\ref{energy}) is of the quadrupole
plus pairing type, with the inclusion of the quadrupole-pairing term
\cite{PSM}
\begin{equation}
\hat H = \hat H_0 - {1 \over 2} \chi \sum_\mu \hat Q^\dagger_\mu
\hat Q^{}_\mu - G_M \hat P^\dagger \hat P - G_Q \sum_\mu \hat
P^\dagger_\mu\hat P^{}_\mu.
\label{hamham}
\end{equation}
This is a rotation-invariant Hamiltonian with each term in it is a
scalar. In Eq. (\ref{hamham}), $\hat H_0$ is the spherical
single-particle Hamiltonian which contains a proper spin-orbit force
\cite{bengtsson85}. The monopole pairing strengths $G_M$ are taken
to be $17.5/A$ for neutrons and $19.5/A$ for protons, where $A$ is
the total particle number. The quadrupole pairing strength $G_Q =
0.16 G_M$, which is the value usually taken in the PSM calculation.
These parameters are found to be appropriate for a correct
description of the known data. To incorporate the shell effect
across a large single-particle space, nucleons in three major shells
($N=3,4,5$ for both neutrons and protons) and in four major shells
($N=4,5,6,7$ for neutrons and $N=3,4,5,6$ for protons) are
activated, respectively, for the smaller deformation region (denoted
as Region I) and for the larger deformation region (denoted as
Region II). The quadrupole-quadrupole interaction strengths $\chi$
are listed in Table I for the regions I and II, and these values are
determined in a similar way as in Ref. \cite{Tanabe00}.

\begin{table}
\caption{Quadrupole-quadrupole interaction strengths (in MeV).}
\label{tab:1}
\begin{tabular}{c|ccc}
 & $\chi_{\rm nn}$ & $\chi_{\rm pp}$ & $\chi_{\rm np}=\chi_{\rm pn}$ \\
\hline\noalign{\smallskip}
Region I & 0.0377 & 0.0302 & 0.0337 \\
Region II & 0.0355 & 0.0284 & 0.0318 \\
\hline\noalign{\smallskip}
\end{tabular}
\end{table}

\section{Detailed analysis of $^{110}$Pd}

\epsfigboxL{fg 1}{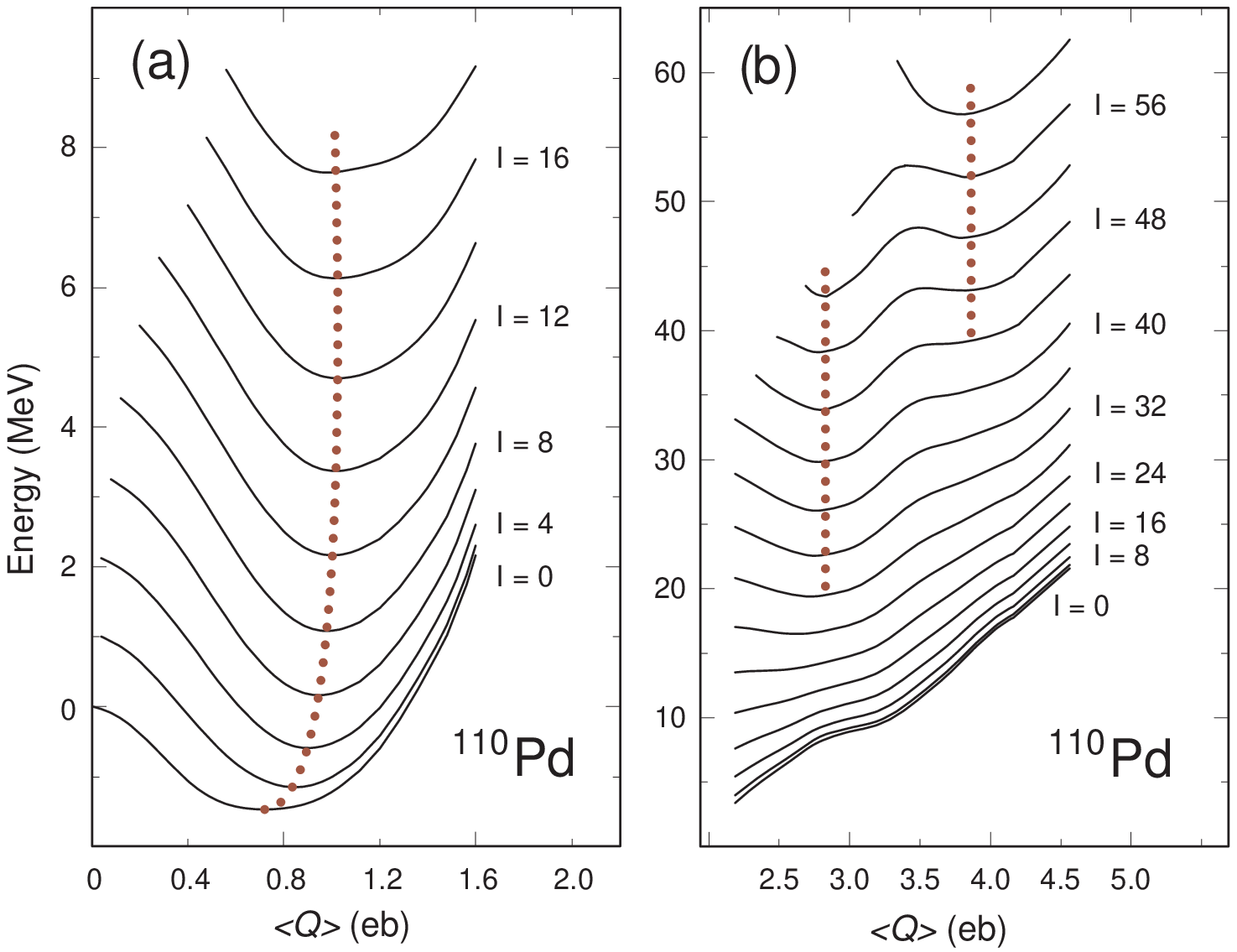}{0pt}{0pt} {Calculated energy surfaces in
$^{110}$Pd for different angular momenta as a function of static
quadrupole moment $\left< Q \right>$. (a) Region I. The energy zero
is set to be the total energy at $\left< Q \right>=\varepsilon_2=0$
and $I=0$. (b) Region II. The absolute energy is arbitrarily chosen.
Note that the plots do not give the relative energy between the two
regions.}

In Fig. 1, we show the calculated energy surfaces for $^{110}$Pd
with different angular momenta. The results clearly indicate three
local minima appearing at distinct deformations (marked by red
dots). The first minimum starts from $\left< Q \right>$ =0.7 eb at
spin $I=0$, and as spin increases, gradually moves to a higher
$\left< Q \right>$ value of 1.0 (corresponding to a deformation of
$\varepsilon_2\approx 0.25$). The change of deformation with spin
indicates a softness of the system at the low angular-momentum
region. At about $I=8$, the shape is stabilized and the system
begins to show a rotor character. We note that Regan {\it et al.}
\cite{Regan03} discussed, with a collection of many experimental
data in this mass region, the evolution of the rotational behavior
in terms of transition from vibrational to rotational motion. Our
present calculation may provide a microscopic insight to the
observation, which suggests that this rotational behavior can be
understood as a gradual increase in deformation with increasing spin
in the first energy minimum \cite{Sun}.

The second minimum is located at about $\left< Q \right>$ =2.8 eb,
corresponding to $\varepsilon_2\approx 0.57$. The minimum is fully
developed only at spin $I=24$ and above. The third one starts
showing up at a rather high spin $I\sim 44$ at about $\left< Q
\right>$ = 3.9 eb, corresponding to $\varepsilon_2\approx 0.79$.
Unlike the first one, the second and the third minimum are stable
({\it i.e.} the deformation minimum does not change with spin),
reflecting a rigid-rotor character. It is worthwhile mentioning that
in a sharp contrast, unprojected energies (not shown in the figure)
suggest a qualitatively different picture: a flat bottom centering
at $\varepsilon_2=0$ and only one rather shallow minimum at
$\varepsilon_2\approx 0.67$. We believe that this difference comes
from the treatment with projection, which incorporates the many-body
correlations missed in the mean-field states.

\epsfigboxS{fg 2}{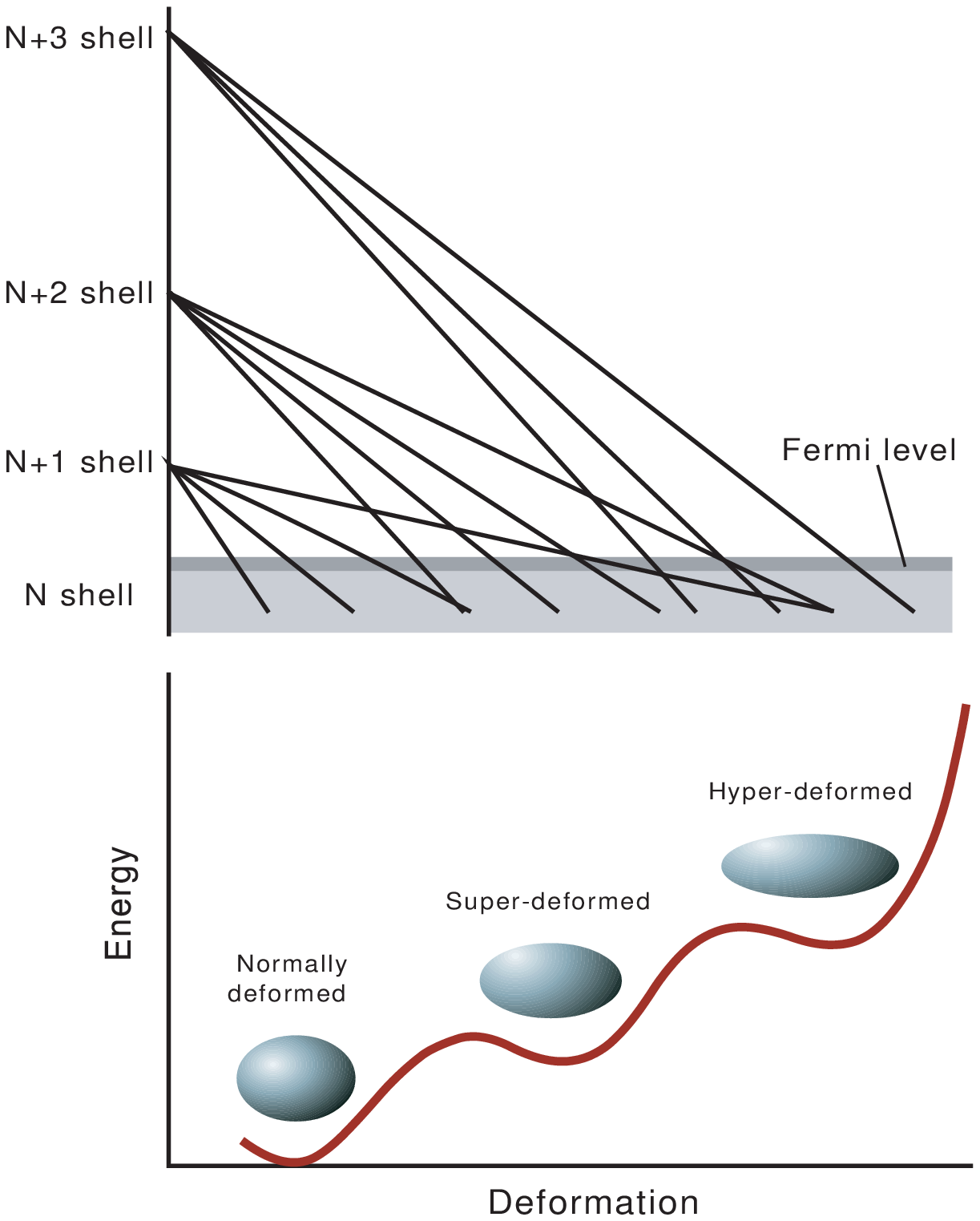}{0pt}{0pt} {Schematic diagram
illustrating a correlation between a formation of deformation minima
and shell mixing near the Fermi level: ND $\leftrightarrow$ adding
$N+1$ shell, SD $\leftrightarrow$ adding $N+2$ shell, and HD
$\leftrightarrow$ adding $N+3$ shell.}

A complete description (including particle-alignment and
back-bending in moment of inertia) of the ND states usually requires
a mixing of the shell with the next $N$-shell (with the high-$j$
intruder orbitals included). Now our calculation indicates clearly
that the SD and HD minima are a consequence of a shell-mixing with
the next two and next three $N$-shells, respectively. This picture
is consistent with the main conclusions of Dudek {\it et al.}
\cite{Dudek87,Dudek04,Schunck07}. As schematically shown in Fig. 2,
the diagram exhibits a correspondence between formation of a
deformed shape and an actual participation of nuclear shells in the
collective motion. It is clear that formation of each higher level
of deformation is associated with particles from an additional major
shell in the shell mixing near the Fermi level. In our example here,
the occurrence of the HD minimum in $^{110}$Pd requires a collective
participation of four neutron shells up to the $N=7$, and four
proton shells up to the $N=6$ shell, respectively.

\epsfigboxM{fg 3}{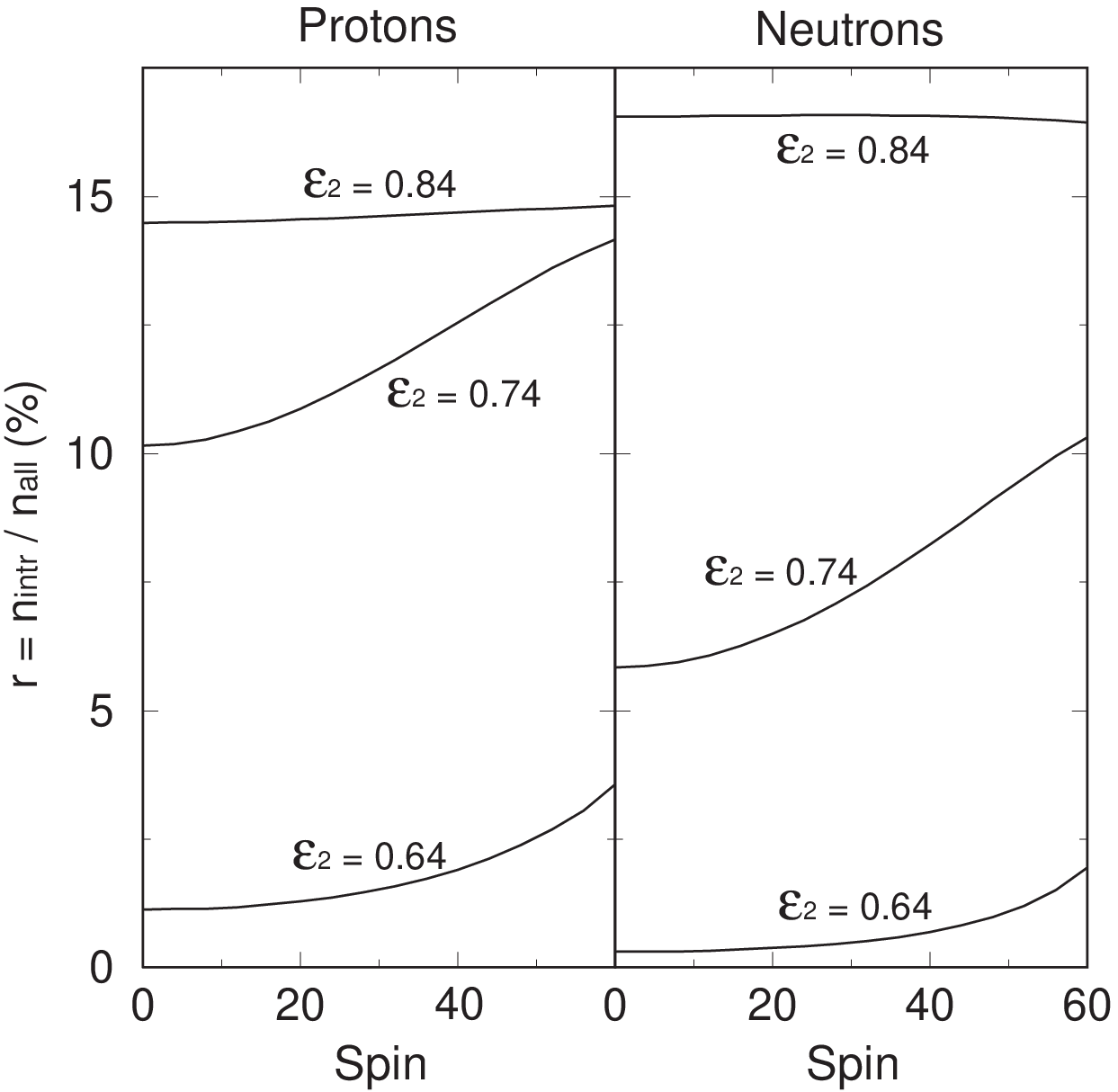}{0pt}{0pt} {Percentage of high-$j$
intruder occupation for different spin states with deformations
around the HD minimum in $^{110}$Pd.}

For a higher $N$-shell, it is evident that among all the orbitals,
the down-sloping intruder orbitals with the highest $j$ are always
occupied first. It is well-known that these intruder orbitals are
important in determining the nuclear high-spin behavior because of
their unique properties with rotation. Thus the next question is
what role these intruder orbitals play in the formation of the
extreme deformation. To answer this question, we consider the
expectation value of the occupation operator
\begin{equation}
n = \left< \Psi^I\right|\hat N = \sum_i c^\dagger_i c_i \left|
\Psi^I\right>. \label{number}
\end{equation}
If the summation runs only over the highest $j$ intruder orbitals we
denote it as $n_{\rm{intr}}$ (for the HD minimum, $j_{{\rm intr}}$
is $j_{15/2}$ for neutrons and $i_{13/2}$ for protons), otherwise as
$n_{\rm{all}}$ when summing over all the orbitals of like particles.
We calculate the ratio $r=n_{\rm{intr}}/n_{\rm{all}}$ for each spin
state by including those orbitals within $\pm 5$ MeV around the
Fermi level.

In Fig. 3, the ratio $r$ obtained at three deformations around the
HD minimum in $^{110}$Pd are separately shown for protons and
neutrons. It can be seen that in both figures, $r$ are rather small
at $\varepsilon_2=0.64$, indicating that occupation of these
intruder orbitals are not important for this deformation. However,
at $\varepsilon_2=0.74$ where the HD minimum is about to develop, a
significant increase in $r$ is observed. In addition, $r$ increase
rapidly with spin, suggesting that the highest $j$ intruder orbitals
play a more important role in high spin states. Finally at
$\varepsilon_2=0.84$ beyond the HD minimum, $r$ reaches the maxima
and stay constant with spin (about $14.5\%$ for the $i_{13/2}$
protons, and $16.5\%$ for the $j_{15/2}$ neutrons). Thus roughly,
the highest $j$ intruder contributes about $1/7$ to the states under
discussion. This is a remarkable value, considering the fact that
this amount comes from only a few intruder orbitals. On the other
hand, the major part of the contribution is indeed from many
particles of the rest orbits although each individual may only
contribute a little.

\epsfigboxM{fg 4}{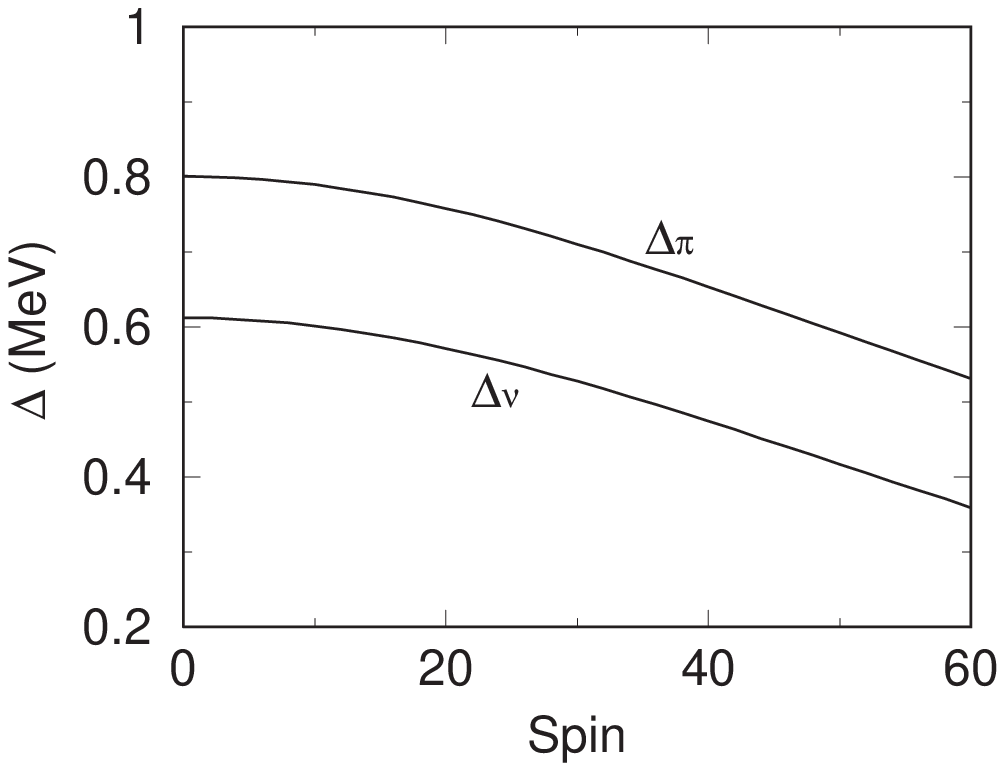}{0pt}{0pt} {Calculated (proton and
neutron) pairing gaps in the HD states in $^{110}$Pd.}

Pairing is one of the most important correlations in nuclear states.
In many high-spin calculations, however, pairing is often neglected
for simplicity. By using the total wavefunction of Eq.
(\ref{wavef}), we have evaluated the pairing correlation in terms of
pairing gaps
\begin{equation}
\Delta_\tau^I = G_M^\tau \left< \Psi^I\right|\hat P_\tau \left|
\Psi^I\right>. \label{pairing}
\end{equation}
In Eq. (\ref{pairing}), $\hat P_\tau$ is the pair operator,
$G_M^\tau$ the pairing strengths in Eq. (\ref{hamham}), and
$\tau=\pi$ or $\nu$. The results for the HD states, with the
corresponding wavefunctions yielding a quadrupole moment $\left< Q
\right>=3.9$ eb, are plotted in Fig. 4. It can be seen that the
pairing gaps decrease with spin as anticipated, which has been
understood as a reduction in pairing correlation due to the
rotational alignment. However, our many-body wavefunction does not
give a sudden collapse in the pair field, or a phase transition from
a superfluid to a normal state. This means that our results suggest
that a considerable amount of pairing correlation remains even in
the highest spin states in the HD well.

\section{Quantitative predictions}

\epsfigboxM{fg 5}{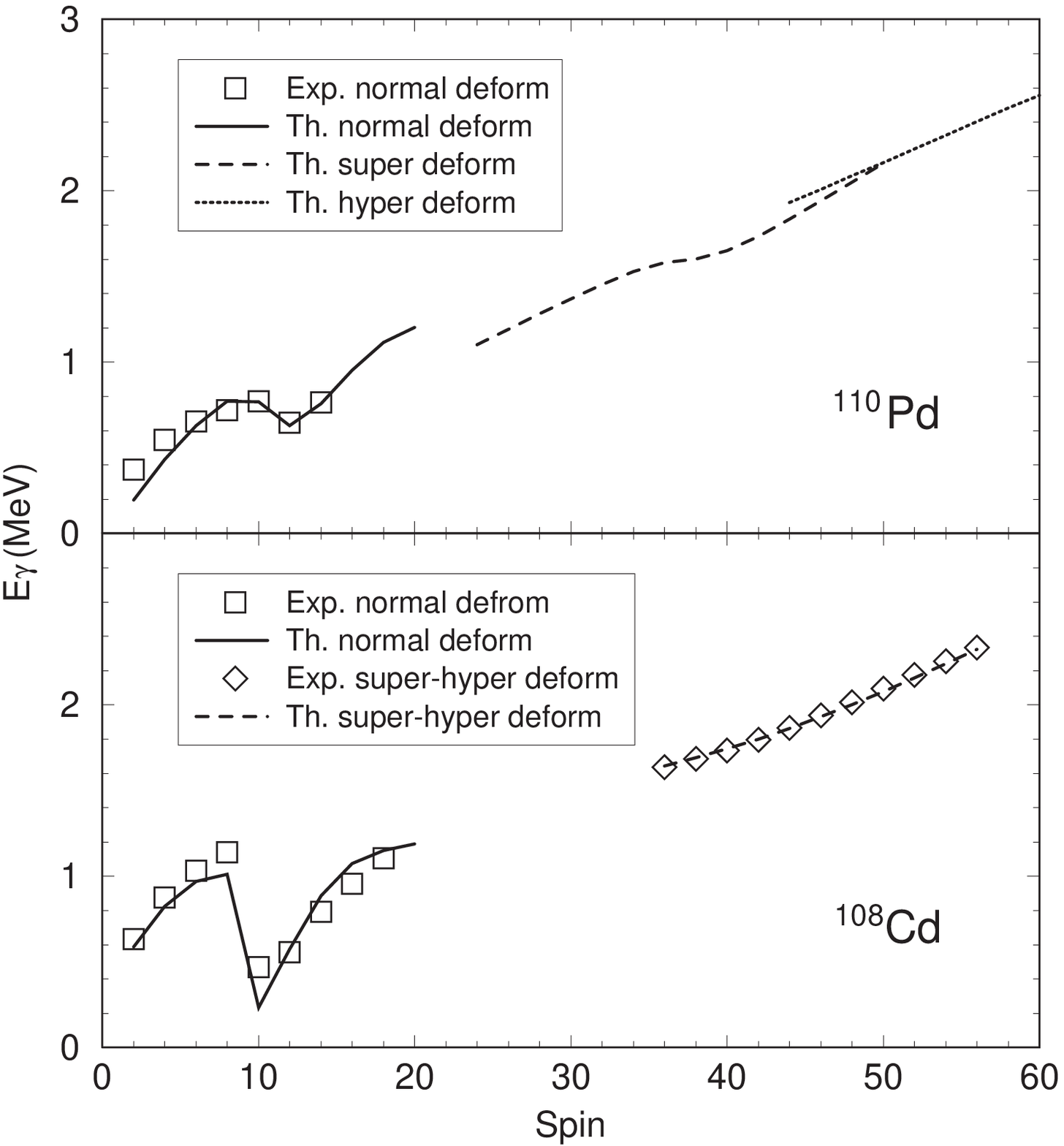}{0pt}{0pt} {Calculated $\gamma$-ray
energies ($E_\gamma(I)=E(I)-E(I-2)$) for $^{110}$Pd (upper panel) at
different energy minima shown in Fig. 1. The theoretical results are
compared with available ND data. Those for SD and HD states in
$^{110}$Pd are predictions. Results for $^{108}$Cd (lower panel) are
also given, and compared with known data.}

We finally present quantitative predictions for yrast energies of
the rotational bands (the lowest state at each angular momentum) in
$^{110}$Pd at each of the minima. In Fig. 5, $\gamma$-ray energies
are calculated by the PSM with the basis deformation
$\varepsilon_2=0.25$ for the ND band, $\varepsilon_2=0.57$ for the
SD band, and $\varepsilon_2=0.79$ for the HD band, respectively. The
experimental ND yrast band in this nucleus is also shown for
comparison. To support the prediction, the PSM results for the two
experimentally-known bands in $^{108}$Cd are also calculated by
using the same calculation condition. As one can see, our results
agree well with the known ND band in $^{110}$Pd, as well as with the
known ND and the highly deformed band in $^{108}$Cd. The SD and HD
bands in $^{110}$Pd are our predictions, awaiting experimental
confirmation.

We have also carried out systematical calculations for the Zr, Mo,
Ru, Pd, and Cd isotopes. Our results indicate that some isotopes of
Zr, Mo, Ru, Pd, and Cd, with the neutron number around 64, show
separate minima corresponding to SD and HD shapes. For all these
nuclei, a SD minimum is found at $\varepsilon_2 \approx 0.56$
starting from $I\approx 24$, and a HD one at $\varepsilon_2 \approx
0.80$ from $I\approx 44$. These predictions are obtained based on a
theory that has quantitatively described the observed band in
$^{108}$Cd \cite{Exp,Lee02}, which, according to our present
results, lies just a step away from the mass region exhibiting
hyper-deformed shapes.

\section{summary}

To summarize, using the projection method implemented in the
Projected Shell Model, we have performed energy surface calculations
for the A-110 mass region. Our results for $^{110}$Pd have indicated
that three distinct energy minima, occurring at different
angular-momentum regions, coexist in one nucleus. We have further
investigated the source for the occurrence of these minima. The
exact treatment of nuclear rotation through angular momentum
projection can be important because the shapes in this study appear
as sensitive functions of angular momentum (e.g. the rapid shape
change in low-spin states at the ND minimum and the emergence of SD
and HD minimum at sufficiently high-spins). Pairing correlation is
sustained in our wavefunctions even at the highest spin of the HD
states. Detailed analysis has shown that, while the nucleons
occupying the high-$j$ intruder orbitals have a larger amount of
individual contributions to the total collectivity, the major part
of collectivity is originated from the rest normal orbitals.

Based on the present study, we have suggested that several nuclei in
the $A$-110 mass region, with the neutron number around 64, would be
the promising candidates in which long-anticipated hyper-deformed
rotational bands may be observed. Thus, the present work has shown a
picture of coexistence of three distinct deformation minima,
corresponding to normal, super-, and hyper-deformed shapes, in
nuclei. We hope that these predictions can be experimentally
confirmed in the near future.

This work was supported partially by the Chinese Major State Basic
Research Development Program through grant 2007CB815005, by U.S. NSF
under contract PHY-0216783 (Y.S.), by U.S. DOE under contract
DE-FG02-96ER40983 (J.-y.Z.), and by NNSF of China under contract
10325521 (G.-L.L.).

\baselineskip = 14pt
\bibliographystyle{unsrt}

\begin{thebibliography} {99}

\bibitem{BMbook} A. Bohr and B. R. Mottelson, {\it Nuclear Structure}
 (W.A. Benjamin, Inc., New York, 1975).

\bibitem{Twin88} P. J. Nolan and P.J. Twin, Ann.\ Rev.\ Nucl.\ Part.
\ Sci.\ {\bf 38} (1988) 533.

\bibitem{Dudek88} J. Dudek, T. Werner, and L. L. Riedinger, Phys.
Lett. B {\bf 211} (1988) 252.

\bibitem{WD95} T. Werner and J. Dudek, At. Data Nucl. Data Tables
 {\bf 59} (1995) 1.

\bibitem{Ch01} R. R. Chasman, Phys. Rev. C {\bf 64} (2001) 024311.

\bibitem{BMP02} B. Buck, A. C. Merchant, and S. M. Perez, Phys.\
Rev.\ C {\bf 65} (2002) 067306.

\bibitem{HD} A. Galindo-Uribarri {\em et al.}, Phys.\ Rev.\ Lett.\ {\bf
71} (1993) 231;\\ D. R. LaFosse {\em et al.}, Phys.\ Rev.\ Lett.\
{\bf 74} (1995) 5186;\\ G. M. Ter-Akopian {\em et al.}, Phys.\
Rev.\ Lett.\ {\bf 77} (1996) 32;\\ A. Krasznahorkay {\em et al.},
Phys.\ Rev.\ Lett.\ {\bf 80} (1998) 2073.

\bibitem{Polon} J. Dudek, N. Schunck, and N. Dubray, Acta Phys. Polon.
{\bf 36}, 975 (2004);\\ P. Fallon, Acta Phys. Polon. {\bf 36},
1003 (2004);\\ H. H\"ubel, Acta Phys. Polon. {\bf 36}, 1015 (2004);\\
B. M. Nyako {\em et al.}, Acta Phys. Polon. {\bf 36}, 1033 (2004).

\bibitem{Exp} R. M. Clark {\em et al.}, Phys.\ Rev.\ Lett.\ {\bf 87}
(2001) 202502.

\bibitem{PSM} K. Hara and Y. Sun, Int. J. Mod. Phys. E {\bf 4} (1995) 637.

\bibitem{Lee02} C.-T. Lee, Y. Sun, J.-y. Zhang, M. Guidry, and C.-L. Wu,
Phys.\ Rev.\ C {\bf 65} (2002) 041301(R).

\bibitem{RS80} P. Ring and P. Schuck, {\it The Nuclear Many-Body Problem}
(Springer, New York, 1980).

\bibitem{Sun95} Y. Sun and M. Guidry,
 Phys. Rev. C {\bf 52} (1995) R2844.
\bibitem{Sun97} Y. Sun, J.-y. Zhang, and M. Guidry,
 Phys. Rev. Lett. {\bf 78} (1997) 2321.
\bibitem{Sun99} Y. Sun, J.-y. Zhang, M. Guidry, and C.-L. Wu,
 Phys. Rev. lett. {\bf 83} (1999) 686.
\bibitem{Sun01} G.-L. Long and Y. Sun,
 Phys. Rev. C {\bf 63} (2001) 021305(R).

\bibitem{HHR84} A. Hayashi, K. Hara, and P. Ring, Phys. Rev. Lett.
{\bf 53} (1984) 337.

\bibitem{Egido93} J. L. Egido, L. M. Robledo, and Y. Sun, Nucl.\
Phys.\ A {\bf 560} (1993) 253.

\bibitem{Egido04} J. L. Egido and L. M. Robledo, Lect.\ Notes\
Phys.\ {\bf 641} (2004) 269.

\bibitem{Ring06} T. Niksic, D. Vretenar, and P. Ring, Phys. Rev. C {\bf
73} (2006) 034308.

\bibitem{Tanabe00} K.-I. Enami, K. Tanabe, N. Yoshinaga,
and K. Higashiyama, Prog. Theor. Phys.\ {\bf 104} (2000) 757.

\bibitem{Valor00} A. Valor, P.-H. Heenen, and P. Bonche,
Nucl.\ Phys.\ A {\bf 671} (2000) 145.

\bibitem{bengtsson85} T. Bengtsson and I. Ragnarsson, Nucl. Phys. A {\bf
436} (1985) 14.

\bibitem{Regan03} P. H. Regan {\em et al.}, Phys.\ Rev.\ Lett.\
{\bf 90} (2003) 152502.

\bibitem{Sun} Details will be published elsewhere.

\bibitem{Dudek87} J. Dudek, W. Nazarewicz, Z. Szymanski, and G. A. Leander,
Phys. Rev. Lett. {\bf 59} (1987) 1405.

\bibitem{Dudek04} J. Dudek, K. Pomorski, N. Schunck, and N. Dubray,
Eur. Phys. J. A {\bf 20} (2004) 15.

\bibitem{Schunck07} N. Schunck, J. Dudek, and B. Herskind,
Phys. Rev. C {\bf 75} (2007) 054304.

\end{thebibliography}

\end{document}